\documentclass[osajnl,twocolumn,showpacs,superscriptaddress,10pt]{revtex4-1} 
\usepackage{amsmath,amssymb,graphicx}

\usepackage{hyperref}
\usepackage{breakurl}
\usepackage{amsmath}
\usepackage{graphicx}
\usepackage{amssymb}

\begin{document}

\title{Nonlinear combining of laser beams}

\author{Pavel M. Lushnikov and Natalia Vladimirova}
\affiliation{Department of Mathematics and Statistics, University of New Mexico, USA}
\email{plushnik@math.unm.edu}


\begin{abstract}
We propose to combine multiple laser beams into a single diffraction-limited beam by the beam self-focusing (collapse) in the Kerr medium.
The beams with the total power above critical are first combined in the near field and then propagated in the optical fiber/waveguide with the Kerr nonlinearity.
Random fluctuations during propagation eventually trigger strong self-focusing event and produce diffraction-limited beam carrying the critical power.
\end{abstract}

\maketitle

The dramatic rise of the output power of fiber lasers in the last 25 years \cite{RichardsonNilssonClarksonJOSAB2011,JaureguiLimpertAndreasTunnermannNaturePhotonics2013} resulted in reaching
$\sim 10$kW in 2009 \cite{GapontsevFominAbramovConf2010} for the diffraction-limited beam. Also 20-kW continuous-wave commercial fiber
laser was announced in 2013  \cite{IPGphotonics} although the beam quality is not yet specified. However, the growth of power since 2009 has been mostly stagnated  because of the encountered mode instabilities
\cite{EidamEtAlOptExpress2011,JaureguiLimpertAndreasTunnermannNaturePhotonics2013}.
The further increase of the total power of the diffraction-limited beam is possible through the coherent beam combining \cite{FanEEEJSelTopicsQuantElec2005,RichardsonNilssonClarksonJOSAB2011} where the phase of each laser beam
is controlled to ideally produce the combined beam with the coherent phase. However, the beam combining has been successfully demonstrated only for several beams.
E.g., Ref. \cite{RedmondEtAlOptLett2012} achieved the combining of five $500$W laser beam into 1.9kW
Gaussian beam with a good beam quality $M^2=1.1$.
Nonlinearity is expected to be the key issue for further scaling of the coherent beam combining \cite{RichardsonNilssonClarksonJOSAB2011}.

Here we propose to use nonlinearity to our advantage to achieve combining of multiple laser beams into a diffraction-limited beam by the strong self-focusing in a waveguide with the Kerr nonlinearity.
The number of laser beams can be arbitrary but we require that the total power to exceed the critical power of self-focusing.
Our estimates below suggest that the commercially available fiber of $\sim 1$mm diameter    \cite{IPGphotonics} might be a possible choice of the waveguide to achieve the diffraction limited beam with the power of several MWs.

We first consider a stationary self-focusing of the laser beam in the Kerr medium
 assuming for now that the pulse duration is long enough to neglect
time-dependent effects. (We estimate the range of allowed
pulse durations below.)
The propagation of a quasi-monochromatic  beam
with a  single polarization through the Kerr media is described by the nonlinear
Schr\"odinger  equation (NLSE) (see e.g.
\cite{BergeSkupinNuterKasparianWolfBergePhysRep2007}):
\begin{eqnarray}\label{nlsdimensionall}
  i\partial_z\psi+\frac{1}{2k}\nabla^2\psi+\frac{kn_{2}}{n_0}|\psi|^{2}
  \psi=
  0,
\end{eqnarray}
where the beam is directed along $z$-axis, ${\bf r}\equiv (x,y)$ are the
transverse coordinates, $\psi({\bf r},z)$ is the envelope of the
electric field,  $\nabla\equiv\left ( \frac{\partial }{\partial x},
\frac{\partial}{\partial y}\right )$,
$k=2\pi n_0/\lambda_0$ is the wavenumber in media, $\lambda_0$
is the vacuum wavelength, $n_0$ is the linear index of refraction, and
$n_2$ is the nonlinear Kerr index.  The index of refraction is
$n=n_0+n_2 I$, where $I=| \psi|^2$ is the light intensity. In fused silica $n_0=1.4535$,  $n_2=3.2\cdot
10^{-16}\text{cm}^2/\text{W}$ for $\lambda_0=790\text{nm}$ and
 $n_0=1.4496$,  $n_2=2.46\cdot
10^{-16}\text{cm}^2/\text{W}$ for $\lambda_0=1070\text{nm}$.

NLSE \eqref{nlsdimensionall} is converted into the dimensionless form
\begin{eqnarray}\label{nls1}
  i\partial_z\psi+\nabla^2\psi+|\psi|^{2}\psi=
  0,
\end{eqnarray}
by the scaling transformation $(x,y)\to (x,y)w_0 $,   $z\to 2zkw_0^2$ and $\psi\to \psi n_0^{1/2}/(2k^2 w_0^2 n_2)^{1/2}$, where $w_0$ is of the order of the  waists of each combined laser beam.

\begin{figure}
\includegraphics[width=0.48\textwidth]{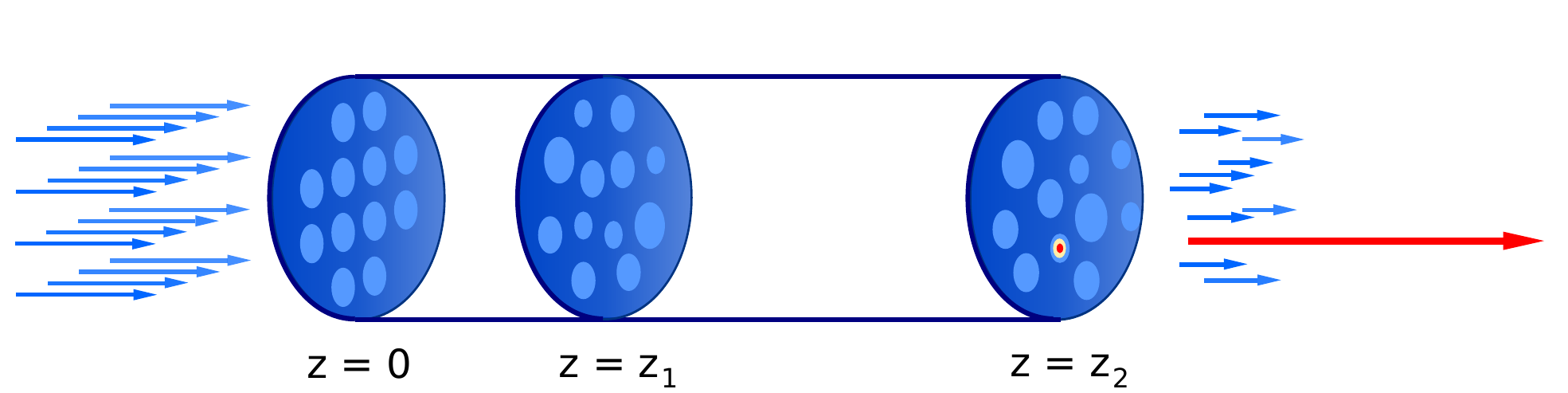}
\caption{(Color online) A schematic of the nonlinear beam
  combining.  An array of beams with non-correlated phases enters a
  nonlinear optical fiber at $z=0$. Inside the fiber the laser field is randomized
  due to nonlinear interactions (see e.g. a schematic of the typical cross-section at $z=z_1$.  A large fluctuation of that random field triggers a strong self-focusing event producing a nearly diffraction-limited hot spot
  at $z=z_2$ (showed by the long arrow) which carriers the critical power $P_c.$  }
\label{fig:nbc_sketch}
\end{figure}

NLSE \eqref{nlsdimensionall} describes the catastrophic self-focusing (collapse) ~\cite{VlasovPetrishchevTalanovRdiofiz1971,ZakharovJETP1972} of the laser beam provided the power $P$ exceeds the critical value
\begin{align} \label{Pcdef}
P_c=\frac{N_c
  \lambda_0^2}{8\pi^2n_2n_0}\simeq\frac{11.70\,\lambda_0^2}{8\pi^2n_2n_0}.
\end{align}
Here
$N_c\equiv 2\pi\int R^2 r dr=11.7008965\ldots$
is the critical power for NLSE \eqref{nls1} in dimensionless units and $R(r)  $ is the radially symmetric Townes soliton \cite{ChiaoGarmireTownesPRL1964} defined as the ground state soliton $\psi=e^{i z}R(r)$  of NLSE with $-R+\nabla ^2 R+R^3=0$, where $r\equiv|{\bf r}|$.
In fused silica $P_c\simeq \text{2MW}$  for $\lambda_0=790\text{nm}$ and  $P_c\simeq \text{4.7MW}$ for $\lambda_0=1070\text{nm}$.

Assume that $N$ laser beams are combined in the near  field (side-by-side combining)
at the entrance $z=0$ to the optical waveguide (the optical fiber) as shown in Fig. \ref{fig:nbc_sketch}. The waveguide can be either multimode optical fiber or any type of waveguide structure
with the Kerr nonlinearity (e.g. it can be a capillar
with the reflective internal walls, filled by a gas or a liquid with the dominated Kerr nonlinearity). We assume that the diameter of waveguide is large enough for the applicability of NLSE \eqref{nls1}. The single polarization is ensured e.g. by the use of the polarization-maintaining optical fiber. We note that a generalization to a  case of arbitrary polazation is possible but  is beyond the scope of this Letter.

The properties of the waveguide in our simulations are taken
into account through the boundary conditions in NLSE along $x$ and $y$. Example is the multimode optical fiber with the diameter in the range between hundreds of $\mu$m to several mm.
At $z=0$ we approximate each  beam to have the Gaussian form with the plane wavefront so that the initial condition for NLSE \eqref{nls1} is the superposition of these Gaussians
$\psi(x,y)|_{z=0} = \sum_{n=1}^N \psi_n$,  $\psi_n = A_n \exp\left(-
\frac{(x-x_n)^2 + (y-y_n)^2}{r_n^2} + i\phi_n\right)$,  where $r_n, \ A_n, \, \phi_n$ and  $(x_n,
y_n) $ are the width, the amplitude, the phase, and the location of the center of  the $n$th beam, respectively. In simulation we assume the same amplitudes $A=A_n$ and widths $r_n=r_0$ for all $N$ beams, but
phases $\phi_n$ are  randomly distributed at $[0,2\pi]$. Randomness of phases $\phi_n$ reflects the randomness in environmental fluctuations and fiber amplifiers of lasers.

\begin{figure}
\includegraphics[width=0.48\textwidth]{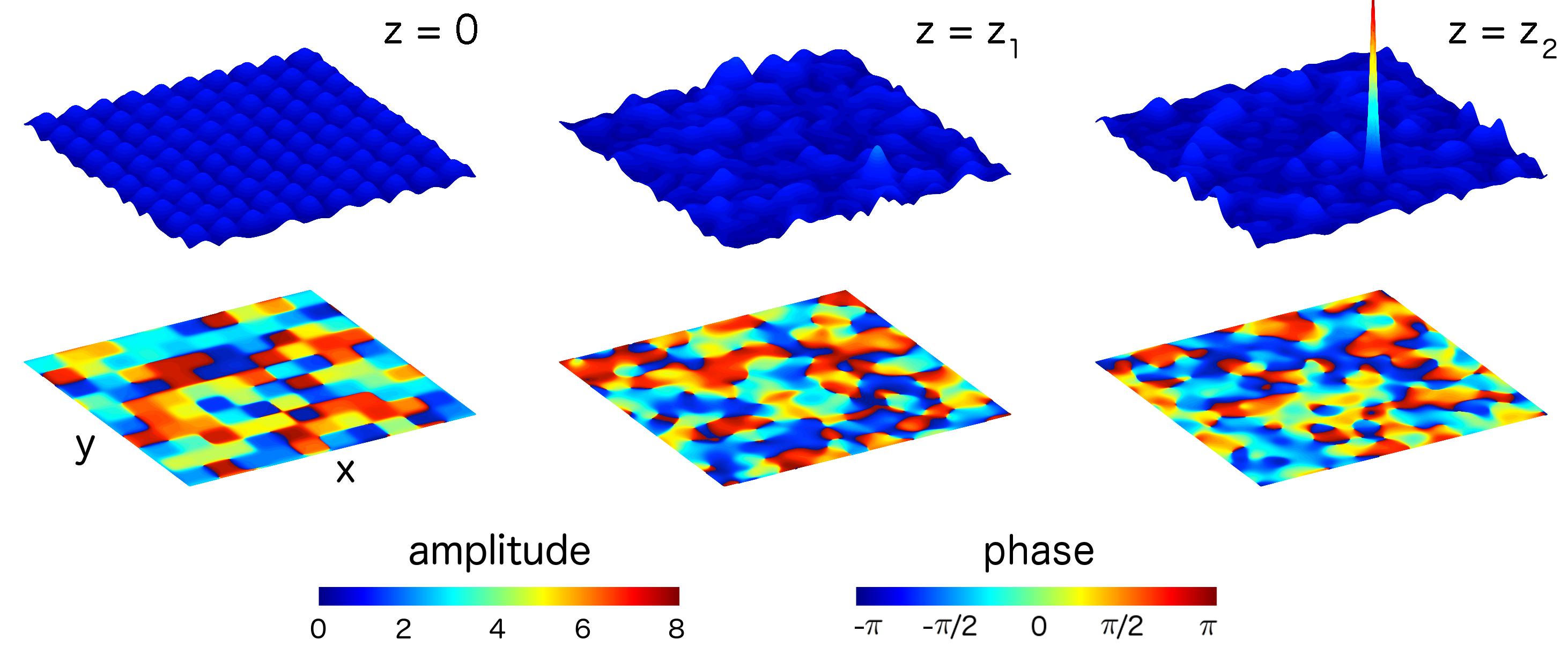}
\caption{(Color online) Simulation of nonlinear beam combining in NLSE (\ref{nls1}).
The snapshots of the distributions (vertical axis) in $(x,y)$ of the amplitude $|\psi|$ (top row) and the phase $arg(\psi)$ (bottom row)  for different values of $z$.
 Left column: the array of Gaussian beams
  with random phases are used as initial conditions ($z=0$).
Middle column: the Kerr nonlinearity results in randomization on phases and
  amplitudes after the propagation distance $z\sim z_{nl}$ as shown for  $z=z_1=10$ ($z_{nl}=5.6$ in that case).
  Right column: the random fluctuations of amplitudes triggers the strong self-focusing collapse event ($z=z_2=15$).}
\label{fig:amp-phase}
\end{figure}

Fig. \ref{fig:amp-phase} shows the typical result of NLSE \eqref{nls1} simulation. We took the square array of $N=10 \times 10$ beams at $z=0$ uniformly located in
the domain $0<x<L, \ 0<y<L, \ L=25.6$. Each beam  had the radius $r_0=1.13$ and carried the power $0.1P_{\rm c}$ (i.e. the total power is $10 P_{\rm c}$).
  A typical
evolution of the system along $z$ is shown in Figure~\ref{fig:amp-phase} for simulations with the periodic boundary conditions in $x$ and $y$.
\begin{figure}
\includegraphics[width=0.48\textwidth]{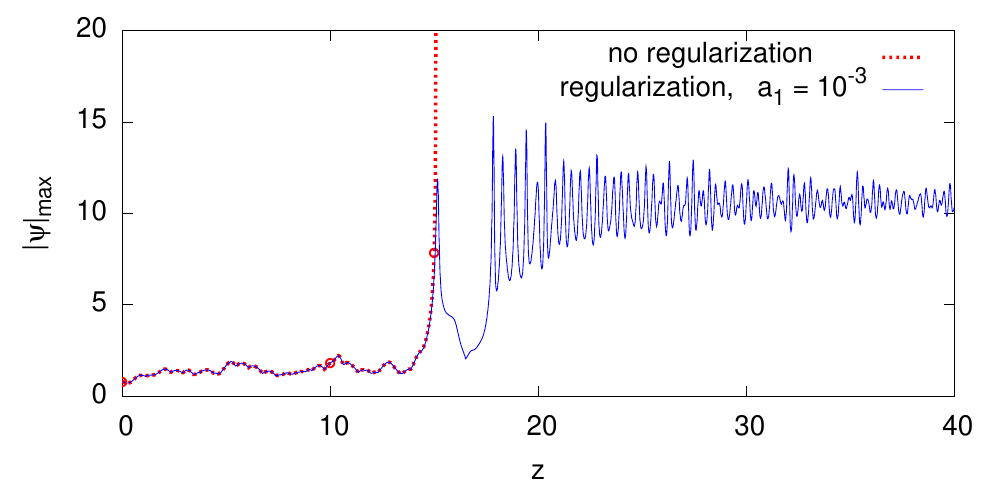}
\caption{(Color online) $\max\limits_{(x,y)} |\psi|$ in the waveguide's
  cross-section vs. $z.$ The dashed line shows for the the result of the same simulation of NLSE \eqref{nls1} as in Fig.  \ref{fig:amp-phase}. The solid line shows the simulation of the regularized NLSE  \eqref{nlssaturated1}
with $a_1= 10^{-3}$ and the same initial condition as for the dashed curve. Thick dots correspond to $z=z_1$ and $z=z_2$ of  Fig.  \ref{fig:amp-phase}.}
\label{fig:pulse}
\end{figure}

\begin{figure}
\includegraphics[width=0.23\textwidth]{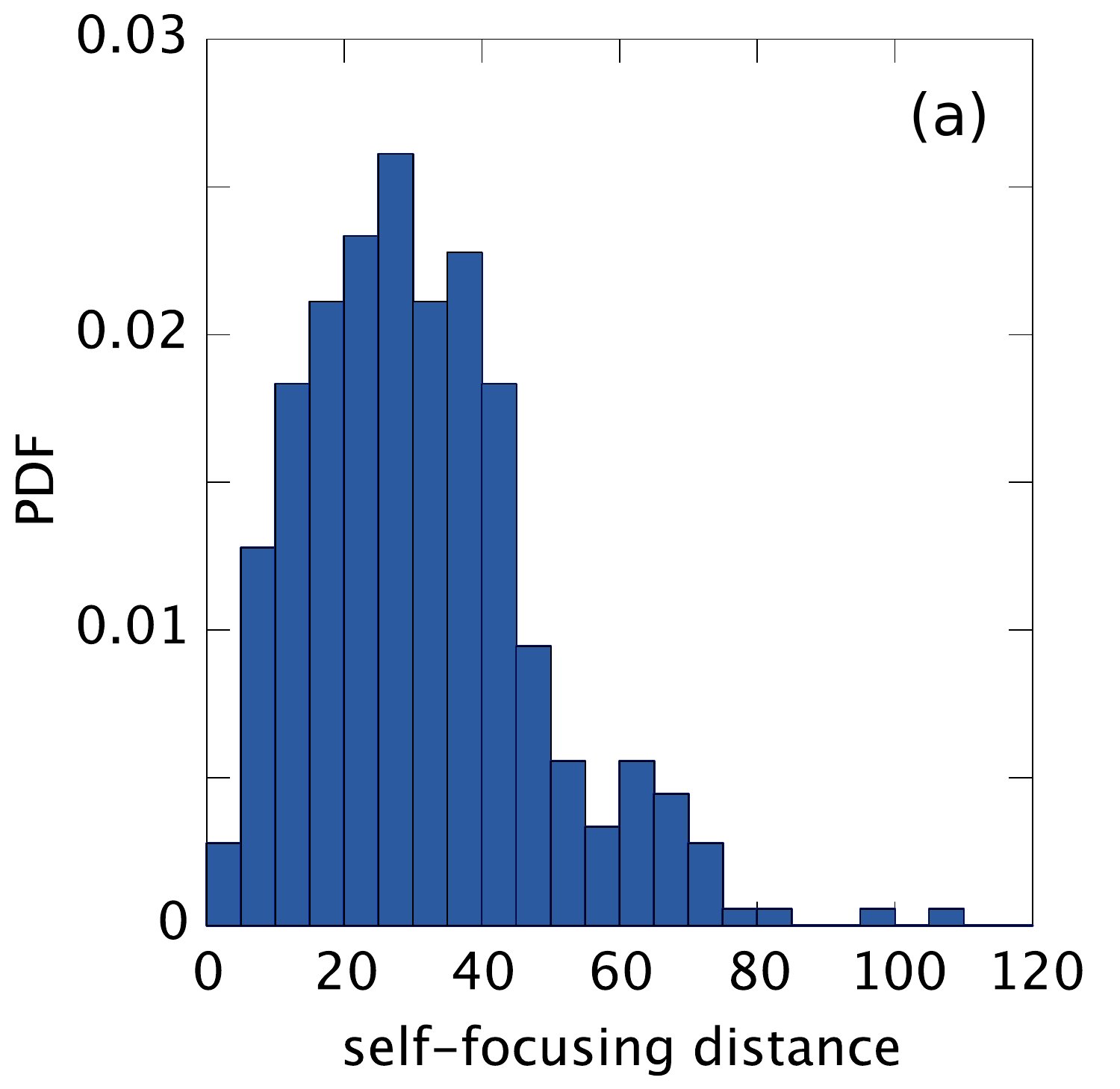}
\includegraphics[width=0.23\textwidth]{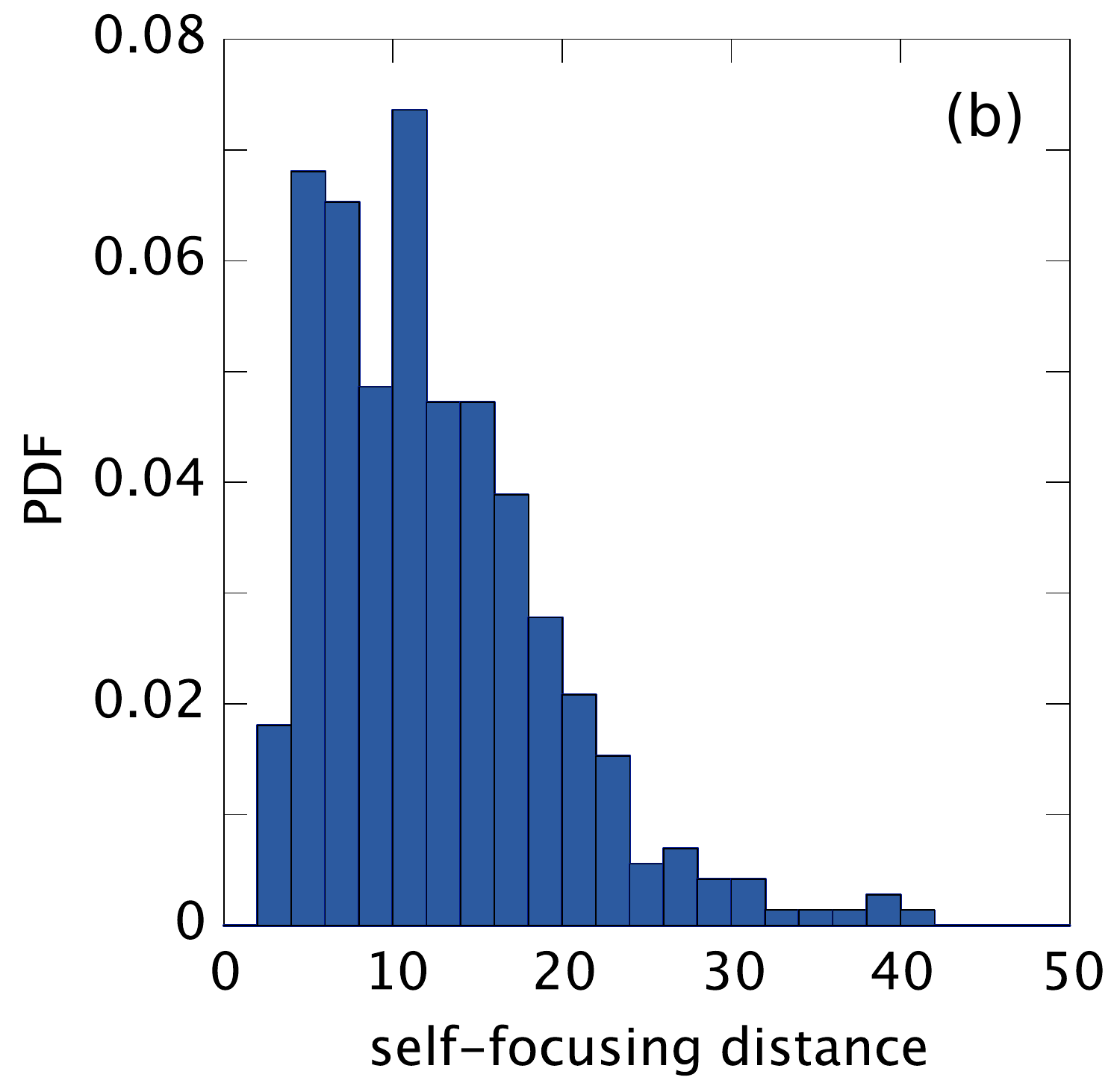}
\caption{(Color online) Probability density functions (PDFs) of
  the catastrophic self-focusing distance $z_{sf}$ collected over 360 simulations with random
  initial phases and the total power  $10 P_{\rm c}$.
  (a) $N=10\times10$ combined beams with $r_0=1.13$.
  (b)  $N=8\times8$ combined beams with $r_0=1.41$.
 }
\label{fig:pdfs}
\end{figure}

The middle column of Fig. \ref{fig:amp-phase} ($z=z_2)$ shows that the amplitudes and phases
become random after propagation of the nonlinear distance $z_{nl}\equiv 1/\langle |\psi|^2 \rangle$, where $\langle |\psi|^2 \rangle=P/S$ is the spatial average of the light intensity in the cross-section area $S$ at
  $z=const$.  For $z>z_{nl}$ the amplitude and phase experience fluctuations along $z$ (optical turbulence) until a large fluctuation    at $z \simeq  15$ in Fig. \ref{fig:amp-phase} triggers strong self-focusing event which results in the formation of large amplitude near diffraction-limited beam (right column of Fig. \ref{fig:amp-phase} shown for $z=z_2$).     These simulations were performed 360
times, for initial conditions with different randomly selected phases
of the input beams.  The probability density function (PDF) of
the distance $z_{sf}$ along the fiber to the point of the first catastrophic self-focusing event are shown in
Fig.~\ref{fig:pdfs}a and \ref{fig:pdfs}b for $10\times 10$ and $8\times 8$ beams, respectively. The power in each beam for these two cases is   $0.1P_{\rm c}$ and  $0.156P_{\rm c}$, respectively. The average value  $\langle z_{sf} \rangle $ (averaged over the ensemble of these 360 simulations)
and the standard deviation  $\langle \delta z_{sf} \rangle\equiv (\langle z_{sf} ^2\rangle-\langle z_{sf} \rangle^2)^{1/2} $ are $\langle z_{sf} \rangle =31.30$,
$\langle \delta z_{sf} \rangle=16.87$ for the simulations of Fig. \ref{fig:pdfs}a and  $\langle z_{sf} \rangle = 12.55$,
$\langle \delta z_{sf} \rangle=6.86$
 for the simulations of Fig. \ref{fig:pdfs}b. We also performed simulations with the added linear potential (circular barrier at $r=0.45L$)
in  \eqref{nls1} to model the boundary of waveguide in transverse directions $(x,y)$ and obtained similar results for PDF (a type of boundary condition is typically essential only for $z<z_{nl}$  provided
the barrier is high enough to make the escape of light from the waveguide a small correction which simulates the total internal reflection).

\begin{figure}
\includegraphics[width=0.48\textwidth]{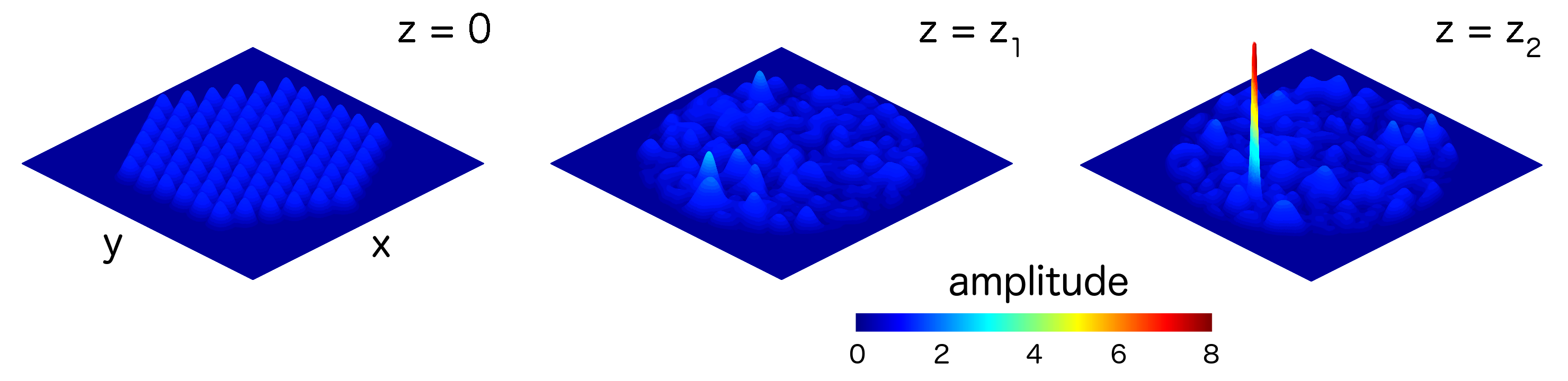}
\caption{(Color online) Simulation similar to Fig. \ref{fig:amp-phase}  but with the added circular barrier at $r=0.45L$ to represent the total internal reflection
of the circular waveguide. 91 beams with the power $0.1P_c$ and $r_0=1.13$ are combined. $z_1=1$ and $z_2=9.7.$
}
\label{fig:amp-phasehexagon}
\end{figure}

The high amplitude beam (the collapsing filament), as in the right row of Fig.  \ref{fig:amp-phase}, is well approximated by the rescaled Townes soliton \cite{SulemSulem1999}:
\begin{eqnarray}\label{selfsimilar}
|\psi(x,y,z)|\simeq\frac{1}{L(z)}R(\rho), \quad \rho\equiv\frac{r}{L(z)}, \quad |{\bf r}|\equiv r,
\end{eqnarray}
where $L(z)$ is the $z$-dependent beam width. The detailed explicit form of $L(z)$ dependence was found in Ref. \cite{LushnikovDyachenkoVladimirovaNLSloglogPRA2013}
starting from the amplitude $|\psi|$ about 3-4 times above the initial value. Thus the collapsing beam of  Fig.  \ref{fig:amp-phase} approaches diffraction-limited beam of the form \eqref{selfsimilar} as it grows in only 3-4 times above the background value $\langle |\psi|^2 \rangle^{1/2}$.
This is also consistent with the study of the optical turbulence dominated by collapses
\cite{LushnikovRosePRL2004,LushnikovRosePlasmPysContrFusion2006,LushnikovVladimirovaOptLett2010,ChungLushnikovPRE2011} that the collapses
are well defined as their amplitudes exceed the background values in 3-4 times.

We also note that  for $z> z_{nl}$ (i.e. after the initial transient propagation), the fluctuations of the intensity   $|\psi|^2$ about $\langle |\psi|^2 \rangle$ have the universal form determined by $\langle |\psi|^2 \rangle$ and $r_0$ \cite{LushnikovVladimirovaOptLett2010,ChungLushnikovPRE2011}. It means that the launching of beams (at $z=0$) with $P>P_c$ into a waveguide unavoidably results in the catastrophic collapse for large enough distance $z_{sf}$ if we neglect waveguide's linear losses as assumed in NLSE  \eqref{nlsdimensionall}. The decrease of the total power closer to $P_c$ only increases $z_{sf}$ (but the value of  $z_{sf}$ always remain finite). Also for very large  $z_{sf}$  one can compensate linear losses by the additional periodical (along $z$) coupling of the waveguide with the external pump.

The regularization of the catastrophic self-focusing depends on the particular type of the Kerr medium. One type of the regularization is the addition of  the saturating nonlinearity
into  NLSE \eqref{nls1}
as follows:\begin{eqnarray}\label{nlssaturated1}
  i\partial_z\psi+\nabla^2\psi+|\psi|^{2}\psi-a_1|\psi|^4\psi=
  0,
\end{eqnarray}
where
$0<a_1\ll 1$.  This type of saturated nonlinearity was found e.g. in chalcogenide glasses with the negative firth order nonlinearity
$n=n_0+n_2 I+n_4I^2$, $n_4<0$  \cite{BoudebsSmektalaEtAlOptCom2003}. The dashed in Fig. \ref{fig:pulse} shows the $z$-dependence of the maximum amplitude $\max\limits_{(x,y)} |\psi|$ for the solution of \eqref{nlssaturated1}
with $a_1= 10^{-3}$ and the same initial condition as for the solid curve of Fig. \ref{fig:pulse}. It is seen that instead of the catastrophic collapse near $z_2=20$ as in NLSE \eqref{nls1}, we observe the periodic
oscillations with the maximum amplitude
 roughly estimated as  $|\psi|\simeq1/a_1^{1/3}$.

Another type of the collapse regularization is the multi-photon absorbtion described by the term  $i\frac{\beta^{(K)}}{2}|\psi|^{2K-2} \psi$ added to the left-hand side (l.h.s.) of NLSE \eqref{nlsdimensionall}. Here
$K$ is the number of photons absorbed by the electron in each
elementary process ($K$-photon absorbtion) and $\beta^{(K)}$ is the
multiphoton absorbtion coefficient. For fused silica with
$\lambda_0=790\text{nm}$ a dominated nonlinear absorbtion
process for this wavelength is $K=5$ with $\beta^{(5)}=1.80\cdot
10^{-51} \text{cm}^{7}\text{W}^{-4}$
\cite{BergeSkupinNuterKasparianWolfBergePhysRep2007} which leads to the formation of plasma and optical damage.

Thus the special measures must  be taken to prevent the damage of the waveguide.
The detailed discussion of that topic is outside the scope of this Letter and we only highlight below several possible ways to overcome that difficulty.
First and perhaps simplest way would be to use the waveguide short enough to avoid a full development of the catastrophic collapse.
Obvious drawback would be that only a fraction of the initial distribution of phases would result in a strong self-focusing producing a near diffraction-limited beam.
Second possible choice is to use a waveguide filled with a gas and ultrashort pulses such that the multiphoton ionization produces plasma which results in the plasma defocusing and clamping
of the collapsing filament.  Such type of clamping has been demonstrated experimentally to allow a formation of filaments of up to several meters in length
\cite{BergeSkupinNuterKasparianWolfBergePhysRep2007} for the propagation of ultrashort pulses in air.
The drawback of that approach is that it would allow beam combining to short pulses only limiting the total energy of the combined beam.
Third option is to use chalcogenide glasses with the negative firth order nonlinearity as described in  Eq. \eqref{nlssaturated1}    \cite{BoudebsSmektalaEtAlOptCom2003}.
Fourth choice is to use of the waveguide with the specially chosen transverse profile of $n_0(x,y)$ and $n_2(x,y)$ such that the collapse starts near the center of the waveguide because of the larger value of $n_0$ there while the catastrophic collapse is stopped by the decrease of $n_2$ is that region \cite{TuritsynPrivate2012}.
Firth choice is nonlinearity management \cite{GabitovLushnikovOL2002} when $n_2$ is periodically modulated
along $z$ to prevent the collapse.
Sixth choice is to form a  ring cavity from the waveguide such that the length of the single round trip along cavity (i.e. along $z$) is not sufficient to achieve catastrophic collapse while the optical switching is used to remove from the cavity the nearly collapsed diffraction-limited beam.
The power depletion from such removal can be compensated by the coupling of the cavity to the laser beams.

To estimate the  parameters for a potential experimental realization of the nonlinear beam combining, we assume that the typical intensity from the combined beams in the waveguide is $I_0=10^9$W/cm$^2$ which allows  continuous-wave (cw)  operation without optical damage
\cite{DawsonALLOptExpr2008}. Consider the  case of  $\langle z_{sf} \rangle =31.30$  for  $10\times 10$ combined beams as in
Fig.~\ref{fig:pdfs}a.
Using the parameters  $n_0=1.4496$, and $n_2=2.46\cdot
10^{-16}\text{cm}^2/\text{W}$ of fused silica at $\lambda_0=1070\text{nm}$ (correspond to the wavelength of the commercially available  50kW cw fiber laser  \cite{IPGphotonics}) we obtain in dimensional units the typical required length of the waveguide
  $l\sim\langle z_{sf} \rangle =4$m and the waveguide thickness $\sim2$mm which is comparable with the commercially available fiber of the 1mm diameter  \cite{IPGphotonics}. Thus we estimate that the combining of several hundred beams from  50kW cw fiber laser  \cite{IPGphotonics} may allow to produce  a nearly diffraction-limited combined beam with the power $\simeq P_c =4.7$MW.  We also note that the high beam quality is not required for each of the combing beams because the self-focusing collapse   spontaneously produces the near diffraction-limited beam from the generic superpositions of combined beams. 

For the pulsed operations, the optical damage threshold is higher than for cw which would allow to achieve nonlinear beam combining in a smaller settings. E.g., typical experimental measurements of the optical damage
threshold in fused silica give the threshold intensity $I_{\rm thresh}\sim 5\cdot
10^{11}\text{W}/\text{cm}^2$ for $8\,\text{ns}$ pulses and $I_{\rm
  thresh}\sim 1.5\cdot 10^{12}\text{W}/\text{cm}^2$ for $14\text{ps}$
pulses \cite{SmithDoApplOpt2008}. Thus the short pulse operations  might allow to scale down the typical lengths $l$ in $z$ and the waveguide cross section in  2-3 orders of magnitude for the same optical power.   However,
for such short pulse durations, $t_0$, we generally might need to take
into account a group velocity dispersion (GVD).  Its contribution is
described by the addition of the term $-\frac{\beta_2}{2}
\frac{\partial^2}{\partial t^2}\tilde\psi$ into the left-hand side of
equation \eqref{nlsdimensionall}.  Here
$\beta_2=370\text{fs}^2/\text{cm}$ is the GVD coefficient for fused silica at
$\lambda_0=790\text{nm}$ and $t$ is the retarded time $t\equiv T-z/c$,
where $T$ is the physical time and $c$ is the speed of light.    At fiber lengths in several meters, the linear absorbtion
of optical grade fused silica is still negligible.  The GVD distance
$\tilde z_{\rm GVD}\equiv 2t_0^2/\beta_2$ must exceed $l$ for
NLSE applicability, which gives $t_0\gtrsim 0.3\text{ps}$ for $l=4$m.

Another possible effects beyond NLSE include a stimulated Brillouin
scattering (can be neglected for the pulse duration $\lesssim
10\text{ns}$ \cite{AgrawalBookNonlinFiberOptics2012} or, similar, if the linewidth of the lasers is made large enough) and a stimulated
Raman scattering (SRS).  The threshold of SRS for a long pulse in
fused silica was estimated from a gain exponent $gI_0l\simeq 16$, where  $g\simeq 10^{-11}\text{cm}/\text{W}$
is the Raman gain constant \cite{AgrawalBookNonlinFiberOptics2012}.
This estimate was obtained assuming that the spontaneous emission is
amplified by SRS (with the amplification factor $e^{gI_0l}=e^{16}$) up to the level of the average light intensity $I_0$ in the waveguide.
 Taking $l=4m$ and   $I_0=10^9$W/cm$^2$  we
obtain the gain exponent $gI_0l\simeq 4\ll
16$,  i.e. we still operate well below the
SRS threshold and can neglect SRS.  This SRS threshold estimate is true for
relatively long pulses $\gtrsim 10\text{ps}$
\cite{AgrawalBookNonlinFiberOptics2012}.  For pulses of shorter
duration, SRS is additionally suppressed because the laser beam
and the SRS wave move with different group velocities.

In conclusion, we demonstrated the possibility to achieve a nonlinear beam combining by propagating multiple laser beams in the waveguide with the Kerr nonlinerity. Large fluctuations  during  propagation seed the collapse event resulting in the formation of near diffraction-limited beam.





\end{document}